\documentclass[reprint,aps,prl,showpacs,floatfix,superscriptaddress,nofootinbib]{revtex4-1}

\usepackage{graphicx}
\usepackage{amsthm}   
\usepackage{amsmath} 
\usepackage{amssymb}  
\usepackage{mathrsfs} 
\usepackage{stmaryrd} 
\usepackage{txfonts} 

\usepackage{enumerate}

\usepackage{appendix}

\newcommand{\E}{{\mathbb{E}}}
\newcommand{\R}{{\mathbb{R}}}

\newcommand{\be}{\begin{equation}}
\newcommand{\bel}[1]{\begin{equation}\label{#1}}
\newcommand{\qe}{\end{equation}}
\newcommand{\ee}{\end{equation}}
\newcommand{\eeq}{\end{equation}}
\newcommand{\ba}{\begin{eqnarray}}
\newcommand{\ea}{\end{eqnarray}}

\date{\today}                      

\begin{document}

\title{Hysteresis effects of changing parameters of noncooperative games}

 \author{David H. Wolpert}
 \affiliation{NASA Ames Research Center, MailStop 269-1, Moffett Field, CA 94035-1000, \tt{david.h.wolpert@nasa.gov}}
 \author{Michael Harr\'{e}}
 \affiliation{Centre for the Mind, The University of Sydney, Australia}
 \author{Eckehard Olbrich}
 \author{Nils Bertschinger}
 \author{Juergen Jost}
 \altaffiliation{Santa Fe Institute, 1399 Hyde Park Road, Santa Fe, NM 87501,
   USA}
 \affiliation{Max Planck Institute for Mathematics in the Sciences, Inselstrasse 22, D-04103 Leipzig, Germany}

\begin{abstract}
We adapt the method used by Jaynes to derive the equilibria of
statistical physics to instead derive equilibria of bounded rational
game theory. We analyze the dependence of these equilibria on the
parameters of the underlying game, focusing on hysteresis effects.  In
particular, we show that by gradually imposing individual-specific tax
rates on the players of the game, and then gradually removing those
taxes, the players move from a poor equilibrium
to one that is better for all of them.

\end{abstract}

\maketitle

\section{Introduction}

The Maximum Entropy (Maxent) principle is an information-theoretic
formalization of Occam's razor. It says that if we are given the
expectation values of some functions of a system's state, then we should
predict that the associated distribution is the one
with minimal information (i.e., maximal entropy)
consistent with those expectations~\cite{coth91,mack03}. Maxent
provides a succinct way to derive much of statistical
physics~\cite{jayn57,jabr03}, e.g., the canonical ensemble.  

Noncooperative game
theory~\cite{futi91,myer91,osru94,auha92} is
the foundation of conventional economics. It uses
provided utility functions of a set of  human ``players" to predict
how the players will model one another. It then uses this to predict
the players' joint behavior.

Many recent applications of statistical physics to economics
analyze it at a coarse-grained level, bypassing its
game-theoretic foundation.
Here we build on~\cite{wolp04a} and apply Maxent to game theory, thereby introducing
statistical physics techniques into the foundation of economics. 

In this application of Maxent, there is a separate expectation
value  for each player. In contrast, when
applying Maxent to derive the canonical ensemble, there is a single
expectation value (of the system's energy).  Accordingly, rather than the
canonical ensemble's single Boltzmann distribution, involving a single Hamiltonian and a
single ``temperature'', we derive a separate
Boltzmann distribution for each player, involving only that player's utility function, and a
``temperature'' unique to that player. The players' Boltzmann distributions are coupled,
and the joint solution provides a bounded rationality version of the Nash equilibrium (NE) of game
theory, where each player's inverse temperature quantifies their rationality.

We analyze the dependence of this modified NE on the
parameters of the underlying game, focusing on bifurcation behavior
and hysteresis effects.  In particular, we show how by gradually imposing taxes on the
players, and then gradually removing them, the joint behavior of the
players can be moved from a poor equilibrium to a Pareto-superior one.
(This can even be done if we require that the players agree to each
infinitesimal change in tax rates, since each such change increases
every player's expected utility.) 
This is particularly interesting given
estimates that non-OECD countries could increase their wealth by one
third by moving from their current equilibrium to a different
one.
 Next we introduce three toy models of how a
society can modify tax rates: via ``socialism'', a
``market'', or ``anarchy''.  We then compare these three models
in terms of the associated discounted sum of total utilities
along the path of tax rates.

\section{Background}

Many different axiomatic arguments establish that the amount  of syntactic information in a
distribution $P(y)$ increases as  the Shannon entropy of that distribution, $S(P) \equiv -\sum_y
P(y) ln[P(y)]$~\cite{jayn57,jabr03,coth91}, decreases.
This provides a way to formalize
``Occam's razor": given limited prior data concerning 
$P(y)$, predict $P(y)$ is the distribution with minimal information (maximum entropy) consistent
with that data.  This formalization of Occam's razor is called the maximum entropy principle
({\bf Maxent}). When the data concerning $P(y)$ is
expectation values of functions under
$P$, Maxent has proven extremely accurate in domains ranging from signal 
processing to supervised learning~\cite{mack03}. Jaynes  used it to
derive statistical physics~\cite{jayn57}, e.g., having
the data be the expected energy of a system or its expected number of particles of
various types.

A finite, strategic form noncooperative game consists of a set of $N$
{\bf{players}}, where each player $i$ has her own set of allowed {\bf{pure
strategies}} $X_i$ of size $|X_i| < \infty$. A {\bf {mixed strategy}} is a distribution $q_i(x_i)$
over $X_i$.  The joint distribution over
$X \equiv \varprod_i X_i$ is $q(x) = \prod_i q_i(x_i)$,
and is called a \textbf{strategy profile}. 

Each player $i$ has
a {\bf{utility function}} $u_i : X \rightarrow \R$.  So given strategy profile $q$, the expected utility of player
$i$ is $\E(u_i) = \sum_x \prod_j q_j(x_j) u_i(x)$
where $q_{-i}(x_{-i}) \equiv \prod_{j \ne i}
q_j(x_j)$.
The {\bf{Nash equilibrium (NE)}} is the strategy profile defined by having every
player $i$  set $q_i$ to maximizes $i$'s expected utility, i.e.,
$\forall i,\ 
q_i = {\mbox{argmax}}_{q'_i}\bigg[ \sum_x q'_i(x_i)  q_{-i}(x_{-i})
u_i(x) \bigg]$.  In general, this set of coupled equations has multiple solutions.

A well-recognized problem of using the NE  to predict real-world behavior
is its assumption that every player chooses their optimal mixed strategy,
which is called {{\bf full rationality}}.
This assumption  is violated (often badly) in many experimental
settings~\cite{came03,star00}. Our modified NE derived using Maxent 
accommodates such {\bf bounded rationality}.

\section{Maxent and Quantal Response Equilibria}
\label{sec:free}

To predict what $q$ the players in a given $N$-player game $\Gamma$ will adopt, first pick one of the
players, $i$. Consider a counter-factual situation, where $i$
has the same move space and utility function as in $\Gamma$, but rather than
have a set of $N-1$ other humans set the distribution over $X_{-i}$, an inanimate stochastic system
sets that distribution, to some $q_{-i}(x_{-i})$.
In general, due to her limited knowledge of $q_{-i}$,
limited computational power, etc., $i$ will choose a suboptimal $q_i$, i.e.,
$q_i \not \in {\mbox{argmax}}_{p_i}[ \E_{p_i q_{-i}}(u_i)]$. 
To quantify this bounded rationality, in analogy to Jaynes' derivation of the canonical ensemble, constrain $q_i$
so that $\E_{q_i, q_{-i}}(u_i)$ has some (nonmaximal)
value for the given $q_{-i}$.
Then Maxent says
\begin{eqnarray}
q_i(x_i) \propto \exp{[\beta_i \E_{q_{-i}}(u_i \mid x_i)]} .
\label{eq:qre}
\end{eqnarray}
where $\beta_i$ is the Lagrange parameter enforcing the constraint.
Note that as $\beta_i
\rightarrow \infty$, $i$ becomes increasingly rational, whereas as
$\beta_i \rightarrow 0$, she becomes increasingly
irrational.

Next, recall that by the axioms of utility theory~\cite{vomo44}, \emph{all}
that player $i$ is concerned with in choosing her mixed strategy is the resultant
expected utility. Accordingly, we presume that if the best
$i$ can do is choose a particular $q_i$ when $q_{-i}$ is set by an inanimate system, she would also choose
$q_i$ if she faces that same distribution $q_{-i}$ when
it is set by other humans. 
%

Generalizing, Maxent says that Eq.~\ref{eq:qre} should hold simulatenously for all $N$ players $i$,
with player-specific Lagrange parameters.
This gives a set of $N$ coupled non-linear equations for $q$. Brouwer's fixed
point theorem~\cite{albo06} guarantees that set always has a solution,
and in general it has more than one.{\footnote{An alternative Maxent
approach would use it to set the entire joint distribution $q(x) =
\prod_i q_i(x_i)$ at once, rather than use it to set each
$q_i$ separately and then impose self-consistency. However there are
difficulties in choosing what constraints to use under this approach.
See~\cite{wolp04a}.}}  

This prediction for $q$
is not based on a model of bounded rational
human behavior derived from experimental data. It is based on desiderata concerning the prediction
process, not on a model of the system being predicted. Nonetheless, it is intriguing to note
that maximizing Shannon entropy has a natural interpretation in terms of a common
model of human bounded rationality, involving the cost of computation. To see this,
recall that $-S(q_i)$ measures the amount of information in the distribution $q_i$.
Say we equate the cost to $i$ of computing $q_i$ with this amount of information.
Then under Maxent, player $i$ minimizes the
cost of computing her mixed strategy, subject to a constraint for the value of 
her expected utility that acts as an  ``aspiration level" . Under this interpretation, $\beta_i$
quantifies $i$'s cost of computing $q_i$, in units of expected utility. 
Future work involves incorporating experimental data concerning human behavior as
additional constraints in the Maxent. (Other
models 
of the cost of computation can be found in~\cite{fule98,hart05,rubi98,rusu95,gepe99}.)

Solutions to our $N$ coupled equations for $q$ are typically called ``logit Quantal Response Equilibria"
(QRE) in game theory~\cite{mcpa95,mcpa96,mcpa08,ango02}. They
have been independently suggested several times as a way to model human
players~\cite{shar04,fukr93,fule98,megi76,ango04,durl99}. 
In all this earlier work the logit distribution is not derived from
first principles.{\footnote{The QRE literature
justifies the logit distribution
by appealing to choice theory~\cite{trai03}, where it arises if
double-exponential noise is added to player utility values. However
that double-exponential noise assumption is never axiomatically justified
in choice theory; it is
assumed for the calculational convenience that it results in the
logit distribution.}} Nor is it  related to information theory,
or the cost of computation. 
Rather typically the logit QRE has been
used as an \emph{ad hoc}, few-degree of freedom model of bounded rational
play. As such it has been widely and successfully used to fit experimental
data concerning human behavior.{\footnote{The logit distribution in Eq.~\ref{eq:qre}
also arises in Reinforcement Learning~\cite{crba96,huwe98a,wotu02a,wotu03a},
as a way to design artificial agents that learn from experience.}}

\section{The shape of the QRE surface}
\label{sec:shape}
\begin{figure}
   \vglue-1cm
	\vglue-20mm
	\hglue-20mm
  	\includegraphics[width=\linewidth]{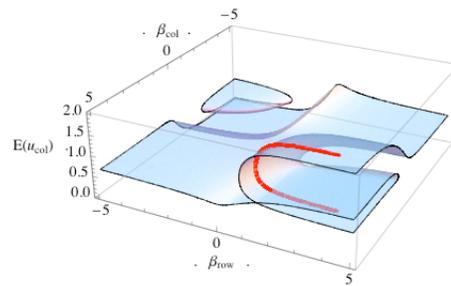}
 	\vglue-47mm
 	\noindent \caption{$\E(u_{col})$ vs. $\vec{\beta}$ under the QRE of the game in Eq.~\ref{table:1}.
         The hysteresis path involving bonds discussed in the text is highlighted.}
 	\label{fig:bigview}
 \end{figure}

To analyze the QRE surface of Eq.~\ref{eq:qre}, we express it as a set of functional relationships,
$q_i=f_i(q_{-i},\beta_i),\quad q_{-i}=f_{-i}(q_{i},\beta_{-i})$. 
A bifurcation may occur if for some $i$
\bel{bif2}
\frac{\partial f_{i}}{\partial q_{-i}}\frac{\partial f_{-i}}{\partial
  q_{i}}\frac{\partial q_{i}}{\partial \beta_{i}}+ \frac{\partial
    f_{i}}{\partial \beta_{i}} -\frac{\partial q_{i}}{\partial
      \beta_{i}} =0
\qe
cannot be solved for $\frac{\partial q_{i}}{\partial
      \beta_{i}}$, i.e.,  if
$\det(\frac{\partial f_{i}}{\partial q_{-i}}\frac{\partial f_{-i}}{\partial
  q_{i}} -\text{Id})=0$.
To illustrate this and related phenomena, we consider games
between a Row and Column player, each
with two pure strategies. The first is the famous ``battle of the sexes"
coordination game~\cite{futi91}, where the  utility functions are
\begin{eqnarray}
2|1 &&\;\;\; 0|0 \nonumber \\
0|0  &&\;\;\;   1|2
\label{table:1}
\end{eqnarray}
where the first (second) entry in each cell is the Row (Column) player's 
utility for the associated pure strategy profile.
 Each joint inverse temperature ${\vec{\beta}} \equiv (\beta_{row},
\beta_{column})$ fixes QRE $q$'s for this game, and therefore 
QRE expected utilities.
Fig.~\ref{fig:bigview} plots this surface taking  $\vec{\beta}$ to $\E_{q}(u_{col})$.
At bifurcations the number of QRE solutions
changes between one and three, and infinitesimal changes in $\vec{\beta}$ may
result in discontinuous changes
in expected utility. (E.g., this happens if the system starts at $\vec{\beta} = (5, 5)$
on the top surface, and then $\beta_{row}$ is reduced to $0$.)
 
\begin{figure}
 	\includegraphics[width=\linewidth]{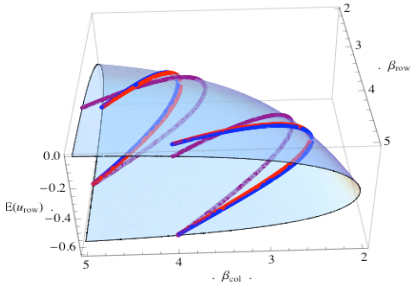}
 	\vglue-10mm
 	\noindent \caption{A QRE surface with paths shown
	for the anarchy (red), socialism (blue) and market (purple) procedures. As
	in Fig.~\ref{fig:bigview}, the x and y axes are player rationalities, $\beta_{row}$
	and $\beta_{col}$, and the z axis is expected utility (this time of player Row).}
 	\label{fig:nicepaths}
\end{figure}

An interesting effect occurs if we multiply the utilities by $-1$.
Fig.~\ref{fig:nicepaths} illustrates part of the surface after this switch.
Note that on the bottom fold, for fixed $\beta_{col}$, decreasing $\beta_{row}$
increases $\E(u_{row})$. So Row benefits by being \emph{less} rational, due to
how Column responds to Row's drop in rationality. 

\begin{figure}
	\includegraphics[width=\linewidth]{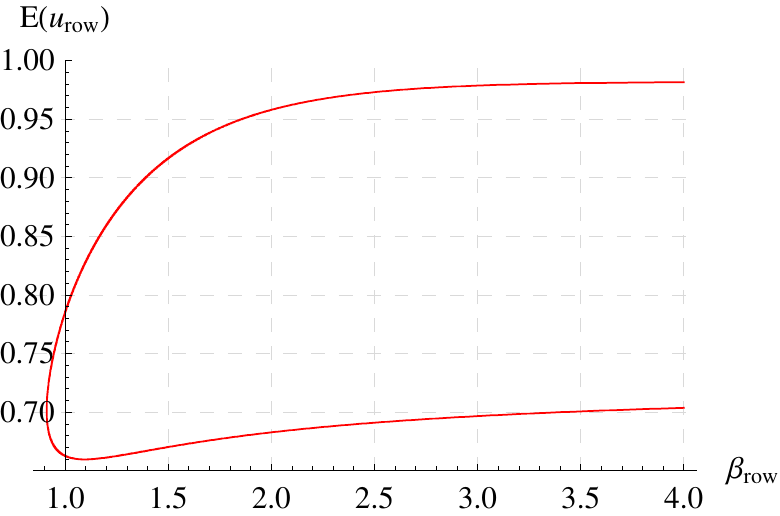}

	\noindent \caption{The expected utility of the Row player along the path through $\vec{\beta}$
	highlighted
	in Fig.~\ref{fig:bigview}, illustrated as a function of the Row player's rationality, $\beta_{row}$.
	The path starts at the bottom right, then travels left, before turning and
	finishing at the top right. Even though the Row player ultimately benefits if society
	follows this path, at the beginning they lose expected utility. They may demand to be compensated 
	for that initial drop, e.g., with proceeds of a bond that are paid off by both players when the end 
	of the path is reached.}
	\label{fig:sub2}
\end{figure}


By Eq.~\ref{eq:qre}, changing $\beta_i$ affects the QRE $q$ the same way as keeping
$\beta_i$ fixed but multiplying $u_i$ by some
$\alpha_i$. So Fig.'s~\ref{fig:bigview},~\ref{fig:nicepaths} give QRE surfaces
where $\vec{\beta}$ is fixed, but each $u_i$ is multiplied by $\alpha_i$.
(Formally, reinterpret the $x$ and $y$ axes as $\alpha_{row}$ and $\alpha_{col}$
rescaled, and reinterpret 
the $z$ axis as $\E_q(u_{i}) / \alpha_{i}$.)
Note we can interpret  $1 - \alpha_i$ as  a tax rate on player $i$. 
So if we model rationalities $\beta_i$ as fixed, e.g., as behavioral attributes, then on the bottom surface in
Fig.~\ref{fig:nicepaths},
Row benefits
if her tax rate \emph{increases}. 

The fact that Row may prefer
a higher tax rate suggests that by varying
tax rates ``adiabatically'' slowly, so that the joint behavior of the
players is always on the QRE surface, we may be able to montonically improve expected utilities
for \emph{both} players. Indeed, by changing tax rates we can
gradually move the equilibrium across the surface from  one fold
to the other, and then undo those changes, returning the rates to their original values, but leaving both 
players with  higher expected utility. (See ~\cite{woku08a}
for other work that exploits the shape of a QRE surface to optimize 
player joint behavior.) 

More precisely, there are paths of $\vec{\beta}$'s (i.e, of $\vec{\alpha}$'s) such that:
\begin{enumerate}
\item Neither player ever
is more rational (taxed at a higher rate) on the path than at the starting point.
\item At each step on the path, if after the next infinitesimal change in $\vec{\beta}$ there is a QRE
$q$ infinitesimally close to the current one, it is adopted. (Adiabaticity.)
\item Each infinitesimal change in $\vec{\beta}$ increases both
$\E_q(u_i)$'s.
\item At each infinitesimal step, if multiple
changes in $q$ meet (1)-(3), but one is
Pareto superior to the others (i.e., better for both players), the players coordinate on that one. 
\end{enumerate}
\noindent Examples of such paths are illustrated in Fig.~\ref{fig:nicepaths}.

The existence of such paths raises the question of how a society should dynamically update its
tax rates. We now compare three procedures for how this could be done
by society as a whole. (For notational simplicity, and to emphasize the analogy with annealing,
we parameterize the procedures in terms of their action on $\vec{\beta}$ rather than on $\vec{1} - \vec{\alpha}$.)
\begin{enumerate}
\item[I. ] 
``Anarchy": Players independently decide how to modify their $\beta$'s.
To do this they follow gradient ascent with a small step size $\Delta$,
subject to the constraint that no player $i$ can go to a $\beta_i$ larger
than the starting one. Thus,
both players $i$ change $\beta_i$ by $\delta \beta_i \in [-\Delta, \Delta]$, using $\partial \E(u_i)/ \partial \beta_i$
to make their choice of what value $\delta \beta_i$ to pick. (Since this is a linear procedure,
the players will always choose one of the three values $\{-\Delta, 0, \Delta\}$.)
\item[II. ] 
``Socialism": An external regulator determines the
path, again using gradient descent, this time  over the sum of the players' expected utilities. At each step of the path $\vec{\beta}$ is changed by
the $(\delta \beta_{row}, \delta \beta_{col})$ vector that maximizes
\be
[\delta \beta_{row} \frac{\partial \E(u_{row})}{\partial \beta_{row}} +
\delta \beta_{col} \frac{\partial \E(u_{row})}{\partial \beta_{col}}]+[\delta \beta_{row} \frac{\partial \E(u_{col})}{\partial \beta_{row}} +
\delta \beta_{col} \frac{\partial \E(u_{col})}{\partial \beta_{col}}]  \nonumber
\qe
subject to $||(\delta \beta_{row}, \delta \beta_{col})||^2 \le
2\Delta^2$. 
(The constraint is to match
the step size to that of the first procedure.)
\item[III. ] 
``Market'':      
Certain mild axioms concerning bargaining behavior of humans
give a unique prediction for what
bargain is reached in any bargaining scenario. Let  $T$ be
the set of joint expected utilities for all the bargains that a set of $N$ bargainers might reach
in a particular bargaining scenario. Then the ``Nash bargaining concept"~\cite{myer91,osru94}
predicts that the the joint expected utility of the bargain reached is  ${\mbox{argmax}}_{{\vec{u}} \in T}[\prod_{i=1}^N u_i]$.

We can use the Nash bargaining concept to predict what change to $\vec{\beta}$
the players would agree to under a ``market" where they bargain with one another
to determine that change. To do this we fix the set of all allowed
bargains to the set of all pairs ${\vec{\beta}}$ such that $||{\vec{\beta}}
- {\vec{\beta}}(t)||^2 \le 2\Delta^2$,
where ${\vec{\beta}}(t)$ is the current joint $\beta$. We also choose $\vec{d}$
to be the joint expected utility at  ${\vec{\beta}}(t)$.
So under Nash bargaining, at each iteration $t$, the players choose the change in joint $\beta$,
$\delta {\vec{\beta}}$, that maximizes the product
\begin{eqnarray}
\bigg[\E(u_{Row} \mid {\vec{\beta}}(t) + \delta {\vec{\beta)}} - 
	\E(u_{Row} \mid {\vec{\beta}}(t)) \bigg]  \;\times \nonumber \\
\bigg[\E(u_{Col} \mid {\vec{\beta}}(t) + \delta {\vec{\beta}}) - 
	\E(u_{Col} \mid {\vec{\beta}}(t)) \bigg] \nonumber
\end{eqnarray}
subject to $||\delta {\vec{\beta}}|| \le 2 \Delta^2$. As in the other two
procedures, we use first order approximations in this one, to evaluate 
the two differences in expected utilities.
\end{enumerate}
In all three procedures the total change in $\vec{\beta}$ in any
step never exceeds $\sqrt{2} \Delta$. This adiabaticity reduces the
computational burden on the players, by not  changing the game too much
from one timestep to the next. (Similar assumptions are called comparitive statics in economics~\cite{keho87}.)

\begin{figure}
 	\includegraphics[width=\linewidth]{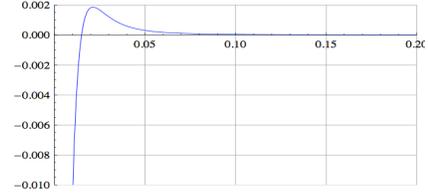}
 	\vglue-70mm
 	\noindent \caption{The difference between the discounted sums of future expected utilities
 	of the two players under the ``socialism" and ``market" procedures, plotted against 
	the discounting factor $\gamma$.}
 	\label{fig:soc_welfare}
 \end{figure}

As in standard economics, we can quantify how good a full path produced
by a procedure is for society as a whole by calculating the discounted
sum of future utilities along the path,
\begin{eqnarray}
Q &\equiv& \sum_{t' > 0} (1 + \gamma)^{t - t'} \sum_{i=1}^N \E(u_i(t'))
\end{eqnarray}
So we can compare  the three procedures by calculating the $Q$'s
for the paths they generate starting from some shared
$\vec{\beta}$ at time $t = 0$. We did this for several representative
initial $\vec{\beta}$'s for the surface in Fig.~\ref{fig:nicepaths}. Anarchy always did
worse than the other two procedures. Those others are compared to each other
in Fig.~\ref{fig:soc_welfare}. When the discounting factor $\gamma$ is large (i.e., we are  more concerned
with near-term  than long-term utility) the market procedure does
better, otherwise socialism does.

All three procedures are local, looking only a single step into the future.
A procedure that also considers the QRE surface's global geometry will produce better
paths in general. In particular, such global information allows us
to consider paths where a player loses expected
utility for certain periods, but in the end all players are better off. 
Fig.~\ref{fig:bigview} highlights such a path, along which player Column always benefits
but player Row loses initially, before ultimately benefitting. (A cross-section of the
expected utility of Row along the
path is shown in Fig.~\ref{fig:sub2}. )
Note that player Row might demand compensation
to agree to follow such a path where they temporarily lose expected
utility,, e.g. in terms of  a subsidy paid for with a bond that 
is repaid by all players at the end of the path. 

Particularly interesting issues arise when setting full
paths under the market model,
if the players use discounted sum of future utilities to value full paths. 
For example, say that at $t=0$ society starts to follow a path $\vec{\beta}_0(t)$ that is a Nash
bargaining solution then. Then in general, for
 $t' > 0$, the path $\vec{\beta}_{t'}(t)$ that is a Nash bargaining solution for full paths
starting from $\vec{\beta}_0(t')$ is not a truncation of $\vec{\beta_0}(t)$ to $t > t'$.  
There is an inconsistency across time. This raises many interesting issues concerning binding commitments,
what it means for a path chosen by bargaining to be renegotiation-proof, etc.

Multiple folds will exist for the QRE surfaces involving many kinds of game parameters, not just tax rates. 
Often such parameters will be set externally, perhaps in a noisy process.
When this is the case,
the QRE surface tells us how stable player behavior is against that external noise. For example,
say the players are on the top fold of the surface in Fig.~\ref{fig:bigview},
with $\vec{\beta} = (2, 4)$,
so the joint behavior is near an edge of
the QRE surface. In this situation, small external noise may lead the players
to ``fall off the edge", and undergo a discontinuous jump to the lower surface.
Moreover, even
if the players managed to (adiababitically slowly) restore their original rationalities after such a jump,
they would end up on the middle fold of the region where $\beta_{row}$ is near $2$,
not on the good fold they started in.
Due to this, when an economic situation exhibits such qualitative
features, it may behoove society to stay away from such edges
in the QRE surface, even if that lowers total expected utility.


%



%

\end{document}